# Electrical switching of Chern insulators in moiré rhombohedral heptalayer graphene

Zhiyu Wang[1#], Qianling Liu[1#], Xiangyan Han[1#], Zhuoxian Li[1], Wenjun Zhao[1], Zhuangzhuang Qu[1], Chunrui Han[2,3*], Kenji Watanabe[4], Takashi Taniguchi[5], Zheng Vitto Han[6,7,8], Sicheng Zhou[9], Bingbing Tong[9], Guangtong Liu[9,10], Li Lu[9,10], Jianpeng Liu[11], Fengcheng Wu[12,13], Jianming Lu[1*]

[1]State Key Laboratory for Mesoscopic Physics, School of Physics, Peking University, Beijing 100871, China

[2]Institute of Microelectronics, Chinese Academy of Sciences, Beijing, 100029, China

[3]University of Chinese Academy of Sciences, Beijing 100049, China

[4]Research Center for Functional Materials, National Institute for Materials Science, 1-1 Namiki, Tsukuba, 305-0044, Japan.

[5]International Center for Materials Nanoarchitectonics, National Institute for Materials Science, 1-1 Namiki, Tsukuba, 305-0044, Japan.

[6]State Key Laboratory of Quantum Optics Technologies and Device, Institute of Opto-Electronics, Shanxi University, Taiyuan 030006, China

[7]Collaborative Innovation Center of Extreme Optics, Shanxi University, Taiyuan 030006, China

[8]Liaoning Academy of Materials, Shenyang, China.

[9]Beijing National Laboratory for Condensed Matter Physics, Institute of Physics, Chinese Academy of Sciences, Beijing 100190, China

[10]Hefei National Laboratory, Hefei, Anhui 230088, China

[11]School of Physical Science and Technology, ShanghaiTech Laboratory for Topological Physics, ShanghaiTech University, Shanghai 201210, China

[12]School of Physics and Technology, Wuhan University, Wuhan 430072, China

[13]Wuhan Institute of Quantum Technology, Wuhan 430206, China

**Abstract**

In orbital Chern insulators, the chemical potential acts as a tuning knob to reverse chirality in dissipationless edge currents, enabling electric-field control of magnetic order—key for future quantum electronics. Despite the rise of orbital Chern insulators, electrically switchable quantum anomalous Hall effect (QAHE) remains rare, necessitating further investigation. Here, we demonstrate electric-field-induced reversal of orbital Chern insulators in a moiré superlattice composed of rhombohedral heptalayer graphene (r-7LG) aligned with hexagonal boron nitride. At one electron per moiré unit cell, two emerging Chern insulating phases—one pointing away from and the other toward graphene's charge neutrality point in the phase diagram of carrier density ($n$) versus magnetic field ($B$)— exhibit energetic competition modulated by both $n$ and $B$. This switchable QAHE chirality in r-7LG demonstrates a layer-number dependent response: similar phenomena in moiré r-6LG require much higher magnetic fields and are absent in r-5LG and r-4LG on the electron side. Our findings establish moiré-engineered rhombohedral graphene as a promising platform for exploring topological quantum materials with electrically controllable chiral edge modes and magnetic order.

## Main

The Chern insulator (ChI) at zero magnetic field exhibits quantum anomalous Hall effect (QAHE), and is characterized by a gapped bulk spectrum and topologically protected chiral edge channels robust against backscattering.[1–3]. This phenomenon provides not only a quantized Hall plateau as a potential metrological resistance standard but also a foundation for ultra-low-dissipation electronics, owing to its non-dissipative edge conduction in the absence of external magnetic fields. Recent advances have extended QAHE functionalities toward electrically reconfigurable chiral states, where reversal of Hall voltage polarity (analogous to switching between binary "0" and "1" states) has been achieved in selected topological platforms through distinct symmetry-breaking mechanisms[4–13]. For example, in Cr-doped $(Bi,Sb)_2Te_3$ heterostructures, spin-orbit torque induced by high-density current pulses enables nonvolatile switching of magnetization, thereby modulating the chirality of edge currents via spin-to-orbital angular momentum transfer[5]. Conversely, in orbital Chern insulators—where intrinsic time-reversal symmetry breaking stems from orbital magnetic moments—electrostatic gating directly manipulates the orbital polarization. This approach circumvents energy-intensive thermal processes, offering a gate-tunable pathway to control chiral edge transport[14].

Prototypical orbital Chern insulators encompass two principal architectures: (i) twisted van der Waals heterostructures[4,7,8,15–24] and (ii) rhombohedral multilayer graphene aligned with hexagonal boron nitride (hBN)[13,25–33] (Figure 1a). In these platforms, flattened moiré bands generate a divergent density of states near the Fermi level, driving Stoner-type ferromagnetism through spontaneous ordering of spin and valley. The valley polarization into time-reversal-symmetry-broken $K/K'$ valleys—hosting opposite Berry curvatures ($\pm\Omega$)—unlocks a finite Chern number ($C$), thereby stabilizing the orbital Chern insulating phase. Crucially, the total orbital magnetization ($M$) comprises two contributions[14,34]: (1) a bulk component arising from wavepacket self-rotation $M_{SR}$, and (2) a boundary term governed by the topological edge current $M_{edge}$, which emerges from electric-field-driven transverse drift of wavepacket. In each valley, when the chemical potential ($\mu$) resides at the Chern gap's lower edge, $M_{SR}$ dominates; As $\mu$ traverses the gap (width $\Delta$), $M_{edge} \propto C \cdot \Delta$ can progressively counteract and even invert the bulk contribution. Because of the time-reversal duality between $K/K'$ valleys, their magnetization are intrinsically antiparallel, enabling synchronized reversal of valley polarization and QAHE chirality via external tuning. As exemplified in Figure 1b, maintaining a fixed direction of $M$ (aligned with $B$) results in toggling electron occupation between $K$ and $K$' valleys hence switching of edge current chirality. Notably, twisted monolayer-bilayer graphene[4] at the filling factor (the number of electron (+)/hole (-) per moiré unit cell) $\nu$ = +3 ($|C|$ = 2) demonstrates this phenomenon, whereas the $\nu$ = +1 state—despite sharing identical $C$—exhibits no such switching, underscoring the critical role of quantum geometry-magnetism interplay. Parallel observations in twisted bilayer graphene systems reveal analogous chirality transitions, albeit with partially quantized Hall plateaus[8–13].

Rhombohedral multilayer graphene (r-NLG) is an emerging material platform for electronic correlation and topological physics. In the moiré superlattice formed by aligning with hBN[13,25–32], rhombohedral graphene exhibits QAHE either in the valence band for r-3LG and r-4LG, or in the conduction band for r-4LG and thicker graphene. Focusing on $\nu$ = +1, i.e., one electron per moiré unit cell, this layer-number dependence can be better identified by how Chern insulators develop[13,25–33]: r-4LG and r-5LG exclusively host $ChI_{C=+1}$ (Here the subscript $C$ is defined for $B$ > 0 throughout the manuscript; Figure 1c); Progressing to r-6LG (Figure 1d), our measurement shows that the coexistence of $ChI_{C=+1}$ and $ChI_{C=-1}$ is enabled—albeit requiring magnetic fields exceeding 0.5 T (See also Ref.[31,33]); In r-7LG (Figure 1e), which is the focus of this work, we find that these competing phases achieve near-degeneracy at ultralow fields (10–20 mT), directly accessing the technologically favorable regime for non-volatile chirality switching. This layer-number tuned topological phase competition constitutes a central discovery of our work. Prior to detailing the electrically reversible transitions between $ChI_{C=+1}$ and $ChI_{C=-1}$, we first map the ground state evolution of the moiré r-7LG system.

**Chern insulators at $\nu$=1 and the nearby Chern wings**

The investigated r-7LG/hBN moiré superlattice exhibits a moiré wavelength of 12.3 nm, calibrated via Brown-Zak quantum oscillations (Extended Data Fig. 3). The displacement field is defined as $D = (C_{bg}V_{bg}-C_{tg}V_{tg})/2$, the carrier density $n = (C_{bg}V_{bg}+C_{tg}V_{tg})/e$, where $C_{tg(bg)}$, $V_{tg(bg)}$, $e$ denote top (bottom) geometric capacitance, top (bottom) gate voltages, elementary charge, respectively. The filling factor is normalized as $\nu = 4n/n_s$, with $n_s$ corresponding to full occupancy of the moiré band (see full $D$-$\nu$ phase diagram in Extended Data Fig. 1). At $D$ < 0 and $n$ > 0, correlated electrons are pushed away from the moiré interface. Figures 2a-b systematically map the longitudinal ($R_{xx}$) and Hall ($R_{xy}$) resistances at 0.5 T, revealing multiple quantized Hall plateaus ($R_{xy} = h/e^2$, blue-shaded regions in Figure 2c). These topological regimes persist at lower fields ($B$ = 0.1T; Extended Data Fig. 2), with device-to-device reproducibility confirmed in Extended Data Fig. 8. While the $v$ = 2 state is dominated by a topologically trivial insulator, the $v$ = 1 state exhibits definitive $|C| = 1$ Chern insulator signatures: quantized Hall resistance of $h/e^2$ and near-zero $R_{xx}$ minima. Moreover, flanking the ν = 1 phase, three wing-shaped regimes (Figure 2c), one on the lower (Wing 1) and two on the higher density side (Wing 2 and 3), display nearly quantized $R_{xy}$ alongside $R_{xx}$ minima.

To gain more insights, we perform magneto-transport spectroscopy at three representative displacement fields: -0.705 (Figure 2d), -0.735 (Figure 2e) and -0.807 V/nm (Figure 2f). For each $D$, the first and second columns respectively present the longitudinal and Hall resistances as functions of magnetic field, with additional $D$-dependent datasets provided in Extended Data Figs. 4-5. Note that the resistance in the range of $B$ = -2 T and 2 T are symmetrized ($R_{xx}$) and anti-symmetrized ($R_{xy}$), whereas for $B$ > 2 T raw data is displayed, demarcated by black horizontal guidelines. The third column is schematics encoding: Chern insulators and wings (blue and red), Landau levels and trivial correlated insulators (grey), and resistance ridges probably from van Hove singularities (purple).

The salient feature is the robust $\mathrm{ChI}_{C=-1}$ at $v$ = 1 persisting across broad $B$–field regimes, which accompanies with (i) a trivial insulator emerging above intermediate fields ($B$ ~0.5-1 T, $D$-dependent) and (ii) a $\mathrm{ChI}_{C=+1}$ dominating above $B$ ~2 T. With increasing $|D|$, the $\mathrm{ChI}_{C=-1}$ is gradually weakened, manifested by the increasing $R_{xx}$ and deviation from $R_{xy} = h/e^2$ quantization at zero fields. Concomitantly, Chern Wing 1 nucleates at $B$ ~0.5 T for $|D|$ > 0.73 V/nm, expanding across the entire $D$-range studied. Crucially, although its $R_{xy}$ approaches $h/e^2$, this wing remains completely isolated from the $v$ = 1 Chern insulator, distinct from extended QAHE which exists over continuous range of carrier densities[30]. For Chern wings 2 and 3, on the one hand they both grow stronger with $|D|$, on the other hand, their competition with electronic phases emanating from $v$ = 1 ($R_{xy}$ > 0, denoted by light red colors) greatly shapes their regimes. A notable feature in Figure 2e is that the Chern wing at $v$ = 1.2 survives at zero $B$ fields, and exhibits a sharp sign reversal along with increasing $v$.

Additional features needing to be mentioned are the incipient Chern phase in finite $B$ fields at $v$ = 2 and resistance ridges (purple-colored curves in Figure 2c-f). The upper ridge in Figure 2c seems not to affect neighboring regimes (Figure 2d-f), whereas the lower ridge is located at the boundary of a hole puddle (Figure 2b). Another hole puddle in the high $B$ field can be found in Extended Data Figs. 4, 5, which forms a closed circle at -0.76 V/nm but grows steadily to the low-$B$ regime at -0.8 V/nm (Figure 2f). Extrapolation of the interior Landau levels indicates that the hole puddle originates from the $v$ = 2 correlated insulator.

**Two competing Chern insulators at $v = 1$**

The above measurements with a coarse field stepping ($\Delta B$ = 50 mT) suggest exclusive dominance of $ChI_{C=-1}$ at low $B$ fields (Figure 2). However, high-resolution field scans ($\Delta B$ = 2 mT) unveil a hidden $ChI_{C=+1}$ phase stabilized at near-zero $B$ fields ($|B| < 5$ mT). Figure 3a maps $R_{xy}$ at $D$ = -0.705 V/nm by fast scanning $v$ (solid arrow) and stepping $B$ (dashed arrow). Obviously, there is a tilted boundary line (white demarcation line) between the two Chern insulators (blue and red, respectively), which is nearly straight. To present a comprehensive phase diagram, we plot the schematic in Figure 3b. Focusing on the quantized Hall region ($v$ = 0.97-1.02, $R_{xy} \sim h/e^2$), we classify the $B$-field evolution into four sectors: (i) Below $|B|$ ~5 mT, the $ChI_{C=+1}$ prevails; (ii) From 5 to 20 mT, the two compete, where the $ChI_{C=+1}$ ($ChI_{C=-1}$) is more favorable at a higher (lower) carrier density; (iii) From 20 mT to 2 T, while the $ChI_{C=-1}$ continues to grow at the lower density side following the Streda trajectory, the $ChI_{C=+1}$ is missed on the higher density side; (iv) above 2 T, $ChI_{C=+1}$ revives and develops following the Streda formula.

We note that the boundary between $ChI_{C=-1}$ and $ChI_{C=+1}$ is quite broad (highlighted by the striped region in Figure 3b), distinct from the abrupt transition at $B$ = 0 T. To quantify the difference (Figure 3c), we extract the section lines in Figure 3a and Extended Data Fig. 6a. Alternatively, we directly scan the $B$ field at $v$ = 0.99 and 1.02 (Figure 3d), and find strong fluctuations in the broad transition region which almost ruin the hysteresis. Indeed, the hysteresis completely disappears under $v$-cycling (Figure 3e).

**Non-volatile electrical switching of orbital orders**

At $D$ = -0.705 V/nm, the coexistence regime of $ChI_{C=-1}$ and $ChI_{C=+1}$ ($B$ = 5-20 mT) exhibits non-hysteretic chirality transitions for $v$ sweeping (Figure 3), indicating that the coexistence is insufficient for non-volatile switching. Upon changing $D$ to -0.735 V/nm, a scan-direction-dependent phase diagram emerges: the forward $v$-sweep enlarges the parameter space of the $ChI_{C=-1}$ (Figure 4a), whereas the backward $v$-sweep expands the space of the $ChI_{C=+1}$ (similar behaviors at $D$ = -0.76V/nm shown in Extended Data Fig. 7). The hysteresis is highlighted in Figure 4c via differential $R_{xy}$ mapping. To better visualize the hysteresis loop, $R_{xy}(B)$ is shown at $B$ = 22, 12, -12 and -20 mT. A surprising observation is the robust switching between two chiral states, regardless whether the electronic state exhibits a well quantized Hall resistance or a merely weak anomalous Hall signal. For example, at -20 mT the resistance (in the unit of $h/e^2$) jumps from +1 (-1) to -1 (+1) during both forward and backward sweeping, respectively. In contrast, at -12 mT the resistance jumps take place when the anomalous Hall effect almost diminishes. If the weak anomalous Hall effect is an indicator of partial valley polarization, this challenges conventional wisdom requiring full valley polarization for hysteric switching of orbital orders.

With these observations, we now turn to the requisite for realizing non-volatile switching in QAHE. First, the competition of two opposite Chern insulators is essential, one of which would point away from ($C$ = +1, assuming electron doping) while the other points towards ($C$ = -1) the charge neutral point of graphene if the Streda trajectories are well developed. In most systems $C$ = +1 prevails, which may arise from light effective mass at higher electron densities[35], i.e., relatively wide renormalized bandwidth. In contrast, $ChI_{C=-1}$ may be destroyed by densely spaced Landau levels due to heavy mass. As this explanation relies on the details in band structures, our layer-number-tuned platform provides an ideal testbed for further theoretical investigation.

Nevertheless, the coexistence at vanishingly small magnetic fields seems not necessary. The electrical switching at 10 mT (Figure 4) resembles ref. [4] and may share the same competition scenario[14]. However, at a small $B$ field, say, 1 mT, we have a definite $ChI_{C=+1}$ by sweeping chemical potential over the full Chern gap, indicating that $M_z$ keeps its positive sign. This is opposite to the case in ref. [4], where the sign of $M_z$ is changed by increasing $\mu$ to a critical value ($\mu_c$) within the Chern gap. At finite $B$ fields, e.g., 10 mT, the competition is restored: following the theory[14], $\mu_c$ now resides in the Chern gap. The mechanism for the dependence of $\mu_c$ on $B$ fields remains to be explored.

Second, the transition boundary in Figure 4 is oblique, distinct from the nearly vertical line in ref. [4]. This means the energy balance of the two states relies on not only $\mu$ but also $B$. To be concrete, we fix the chemical potential, i.e., carrier filling $\nu$ in Fig. 3a at $\nu$ = 1, and sweep $B$. At $B$ = 0 T, the two Chern states are degenerate. When increasing up to 8 mT, $ChI_{C=+1}$ is observed because of its lower energy originating from the coupling between its positive magnetic moment and $B$ fields. However, when further increasing to 15 mT to cross the transition boundary, $ChI_{C=+1}$ will exhibit a higher energy. Consequently, one can infer that the $B$ dependence of the energy competition is non-monotonic. Considering the definition of magnetization, $M \equiv -\partial F/\partial B$ with $F$ being the free energy, the magnetic moment appears to change with the $B$ field.

At last, we notice that our observation resembles the competing Chern insulators in twisted bilayer-trilayer graphene[6] and moiré rhombohedral tetralayer graphene (one hole per moiré unit cell)[27], where the transition boundary can be tuned by an in-plane $B$ field. Whether the competition in our sample also relies on in-plane $B$ fields deserves to be investigated in the future. Another interesting observation is the similar layer-number dependence on the hole side of moiré rhombohedral multilayer graphene: for one hole per moiré unit cell, only $ChI_{C=-1}$ is observed in r-3LG[25,26], whereas in r-4LG $ChI_{C=+1}$ also exists (although with the assistance of a small in-plane $B$ field)[27]. Despite asymmetric performances of electrons and holes, the argument that coexistence of $ChI_{C=-1}$ and $ChI_{C=+1}$ prefers thicker rhombohedral graphene remains valid.

**Conclusion**

We demonstrate nonvolatile chirality control in the quantum anomalous Hall effect of moiré-engineered rhombohedral heptalayer graphene. The electric-field-induced chirality switching enables concurrent modulation of three key parameters: (i) Hall voltage polarity, (ii) dissipationless edge current direction, and (iii) magnetic order. This tripartite manipulation establishes a unique platform for developing hybrid memory-logic architectures with configurable electrical/magnetic operation modes. Considering anticipated progress in high-temperature quantum anomalous Hall systems, our findings provide critical insights for implementing orbital Chern insulators in next-generation computing paradigms, particularly for energy-efficient compute-in-memory technologies requiring nonvolatile state control[37,38].

## Acknowledgements

This work was supported by National Key R&D Program of China (grant no. 2021YFA1400100 (J.L.), 2024YFA1409700 (J.L.), 2022YFA140240 (F.W.)), NSF of China (grant no. 12374168 (J.L.), 62275265 (C.H.)), Beijing Natural Science Foundation (grant no. 4222084 (C.H.)). This work was supported by the Synergetic Extreme Condition User Facility (SECUF, https://cstr.cn/31123.02.SECUF). We also acknowledge the support from Peking Nanofab (J.L.).

## Contributions

J.L. and C.H. conceived the project. Z.W., Q.L. and X.H. fabricated devices with assistance from Z.L., W.Z., Z.Q., and performed transport measurements with assistance from B.T. and G.L.; K.W. and T.T. synthesized boron nitride crystals; J.L., Z.H. and C.H. supervised the project. J.L., F.W. and J.P.L. performed theoretical analysis. All authors contribute to the data analysis. Z.W., Q.L., X.H. and J.L. wrote the paper with input from all authors.

## Competing interests

The authors declare no competing interests.

## Figures

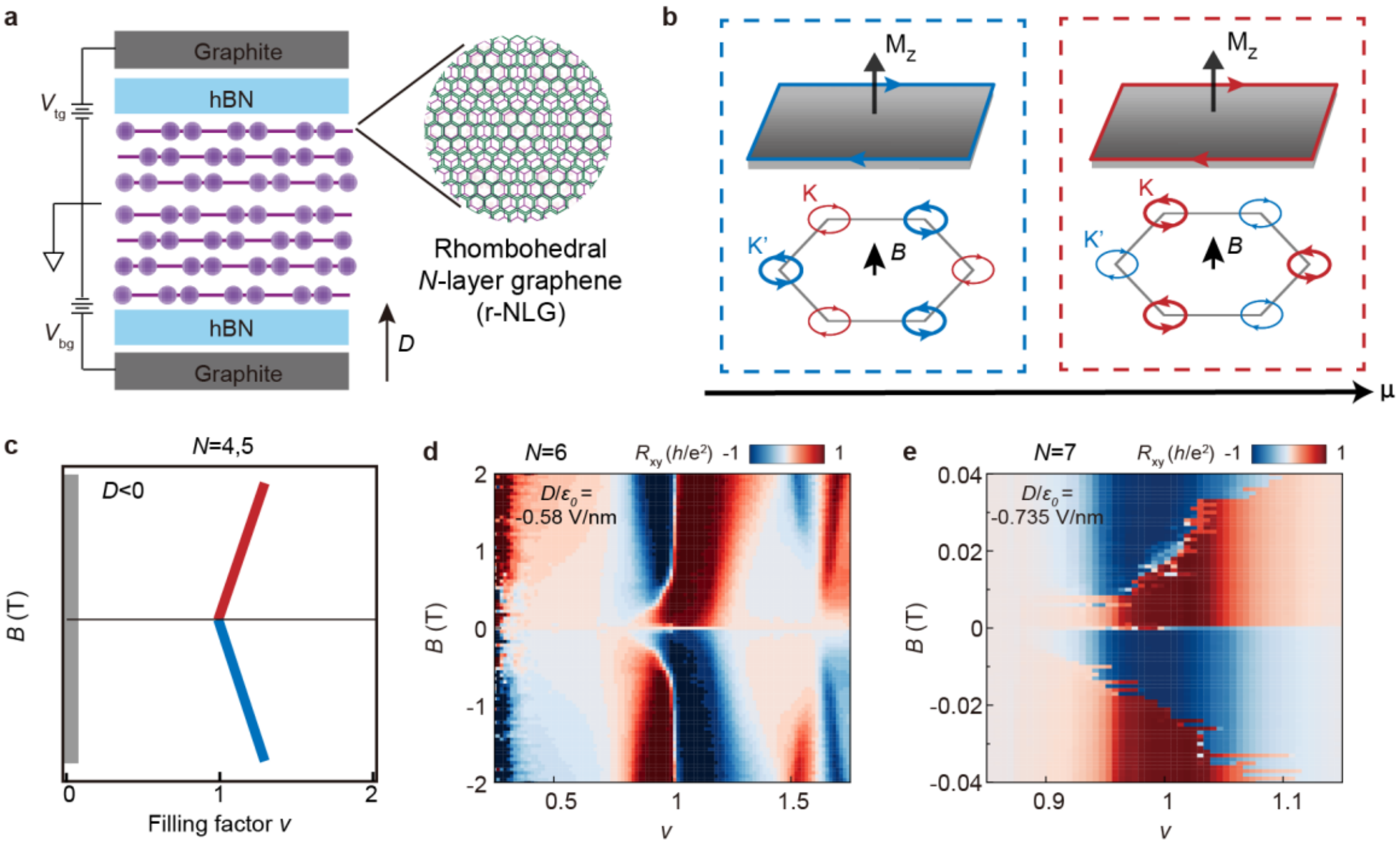

**Figure 1 Evolving competition of Chern insulators at $\nu$ = 1 in rhombohedral multilayer graphene. a,** Device schematic. The moiré heterostructure comprises rhombohedral *N*-layer graphene (r-NLG) encapsulated between top/bottom hexagonal boron nitride flakes (h-BN), with vertical displacement field *D* applied between back ($V_{bg}$) and top gate ($V_{tg}$). **b,** Valley-selective magnetization reversal. As chemical potential $\mu$ traverses the Chern gap, orbital magnetization $M_Z$ in *K'* valley (positive initial state) and *K* valley (negative initial state) undergoes sign inversion due to competing bulk-edge contributions. This valley polarization switching under positive magnetic field *B* induces concomitant reversal of both Hall voltage polarity and edge current chirality. **c-e,** Layer-number-dependent Chern phase competition. *N* = 4, 5 systems exhibit unidirectional Chern insulation away from charge neutrality point (CNP) across $\nu$-*B* phase space (**c**). *N* = 6 introduces competing CNP-directed Chern states emerging at *B* ~0.5 T (**d**, moiré wavelength ~ 13.1 nm). *N* = 7 enables low-field (~10 mT) phase competition (**e**).

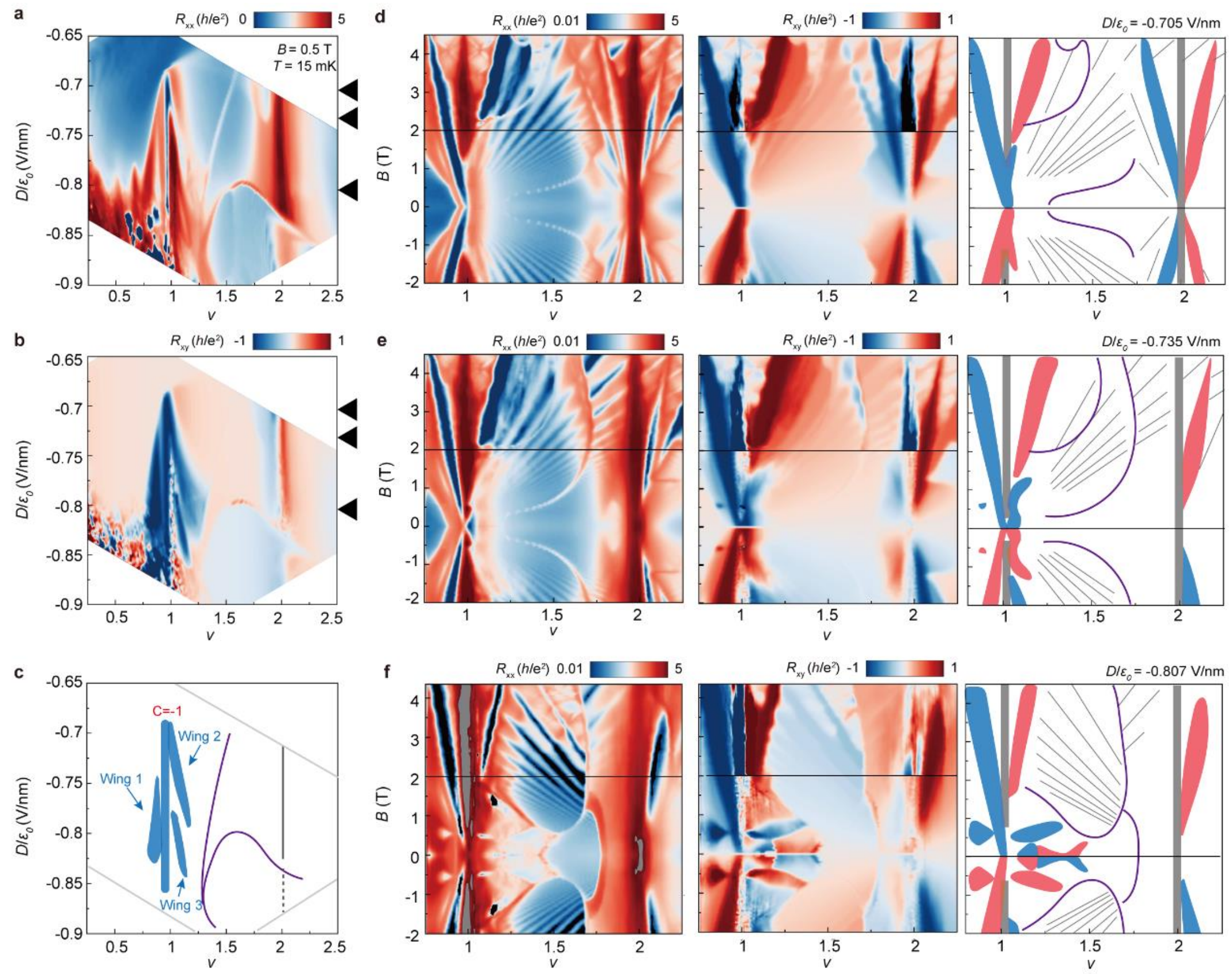

**Figure 2 Chern insulators at $\nu$ = 1 and proximal wings as a function of displacement fields. a-b,** $D$-$\nu$ phase diagrams of longitudinal (**a**) and Hall (**b**) resistance taken at $B$ = 0.5 T and $T$ = 15 mK. Contrasting with correlated resistive peaks at $\nu$ = 2, the $\nu$ = 1 regime exhibits a well-developed Chern insulator and three wings on two sides. Solid triangles denote selected $D$s for panels **d-f**. **c,** Schematic of main features: Chern insulating phases (blue), resistive ridges (purple), and trivial correlated insulator (grey). **d-f,** Field-dependent transport response at $D$ = -0.705 V/nm (**d**), -0.735 V/nm (**e**), and -0.807 V/nm (**f**), showing $R_{xx}$ (left, symmetrized), $R_{xy}$ (middle, anti-symmetrized), and corresponding schematics (right). Raw data ($|B|$ > 2 T) and processed data ($|B| \leq 2$ T) are demarcated by black horizontal lines. Color-coding scheme: Chern insulators (blue/red), Landau levels and trivial correlated insulator (gray), and resistive ridges (purple).

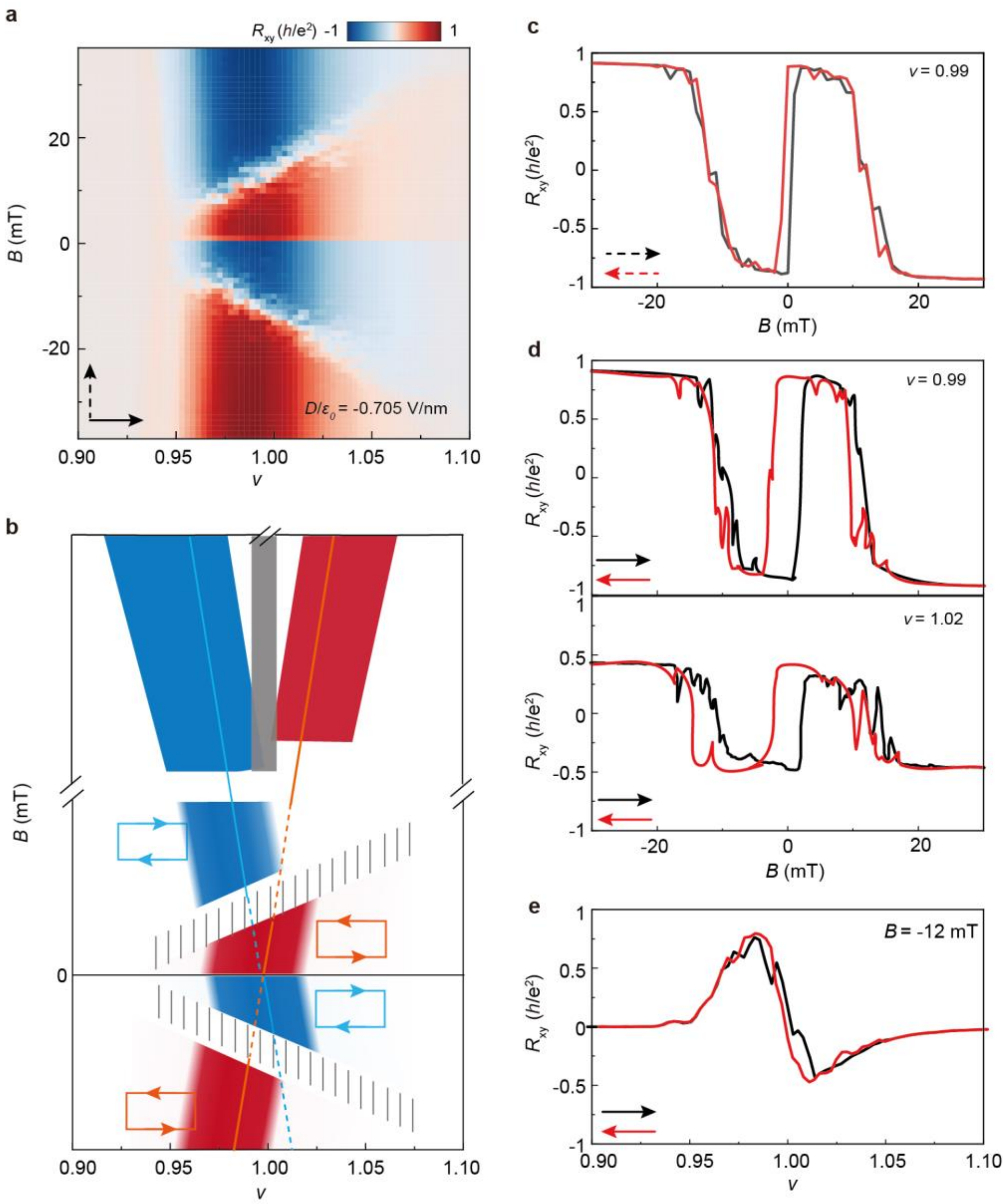

**Figure 3 Competing orbital Chern chiralities at $D$ = -0.705 V/nm. a,** High-resolution mapping via $\nu$-sweep (fast axis, denoted by a solid arrow) and $B$-stepping ($\Delta B$ = 2 mT, slow axis, denoted by a dashed arrow) reveals $ChI_{C=-1}$/$ChI_{C=+1}$ competition. The zero-field regime shows $B$-polarity-locked chirality ($C$ = +1 at $B$ > 0; $C$ = -1 at $B$ < 0), which undergoes polarity inversion beyond critical field that strongly depends on $\nu$. **b,** Unified phase transition schematic integrating Fig. 2d and Fig. 3a. Red/blue trajectories obey Streda formula ($C$ = ±1) with transition boundary (denoted by strips). **c,** Section lines from forward (panel a) and backward (Extended Data Fig. 6b) $B$-stepping demonstrate stark contrast: abrupt phase transition at $B$ = 0 T vs. broad and continuous transition in the stripped region in panel b. **d,** Hysteretic response shows strong bistability near $B$ = 0 versus fluctuation-dominated secondary transition. Top: $\nu$ = 1; bottom: $\nu$ = 1.02. **e,** Bidirectional $\nu$ sweeps show negligible hysteresis.

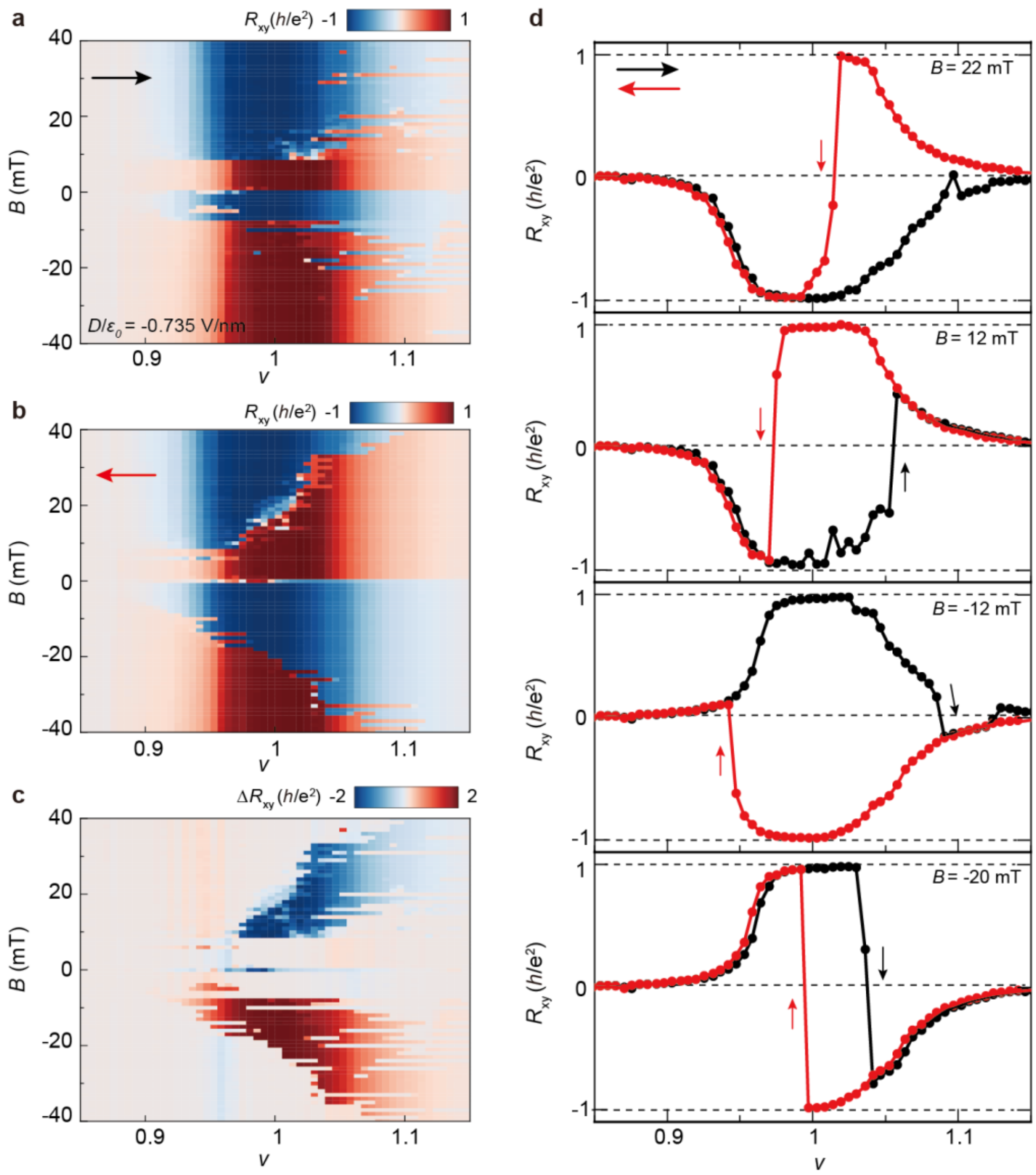

**Figure 4 Nonvolatile orbital order switching via electric-field control at $D$ = -0.735 V/nm. a-c,** Hysteresis mapping: (**a**) forward $\nu$-scan, (**b**) reverse $\nu$-scan, and (**c**) their difference revealing bidirectional orbital polarization reversal. **d,** $B$-field-modulated hysteretic memory showing chirality reversal (top to bottom: $B$ = +22, +12, -12, -20 mT). At switching points (highlighted by vertical arrows), the states are either fully valley-polarized (bottom) or merely partially polarized.

## Methods

### Sample fabrication

The crystals used in this study were natural graphite, which were mechanically exfoliated to obtain thin layers on $SiO_2$ (285 nm)/Si substrates. The number of graphene layers was determined by optical contrast, while the stacking order was assessed through near-infrared imaging. The thickness of the boron nitride was measured using atomic force microscopy (AFM). The transfer process followed the standard dry transfer method. To increase the success rate of transferring rhombohedral graphene, we employed a 532 nm picosecond pulsed laser to cut the graphene into independent rhombohedral stacking regions. During transfer, we carefully aligned the straight edges of the boron nitride with those of the graphene to form a moiré superlattice. Subsequent fabrication steps included electron beam lithography, inductively coupled plasma etching, and thermal evaporation of Cr (1 nm)/Au (50 nm) to finalize the device.

### Electrical measurement

Low-temperature electronic transport measurements were conducted using a dilution refrigerator (Oxford Kelvinox) with a base temperature of 15 mK. A standard low-frequency lock-in technique (LI5650) operating at 17.7777 Hz was used to measure the longitudinal and Hall resistances of the Hall bar devices. The alternating excitation current was set as 0.5 nA. Gate voltages were applied using Keithley 2612 and Keithley 2400 instruments.

To minimize the potential interference of longitudinal resistance in the Hall resistance measurements, the standard symmetrization method was applied, where for resistance measured at a fixed magnetic field $B$, $R_{xy}(\pm B) = \frac{R_{xy}(B)-R_{xy}(-B)}{2}$ and $R_{xx}(\pm B) = \frac{R_{xx}(B)+R_{xx}(-B)}{2}$.

## Data availability

The data that support the findings of this study are available within the article and Supplementary Information. Any other relevant data are available from the corresponding authors upon reasonable request.

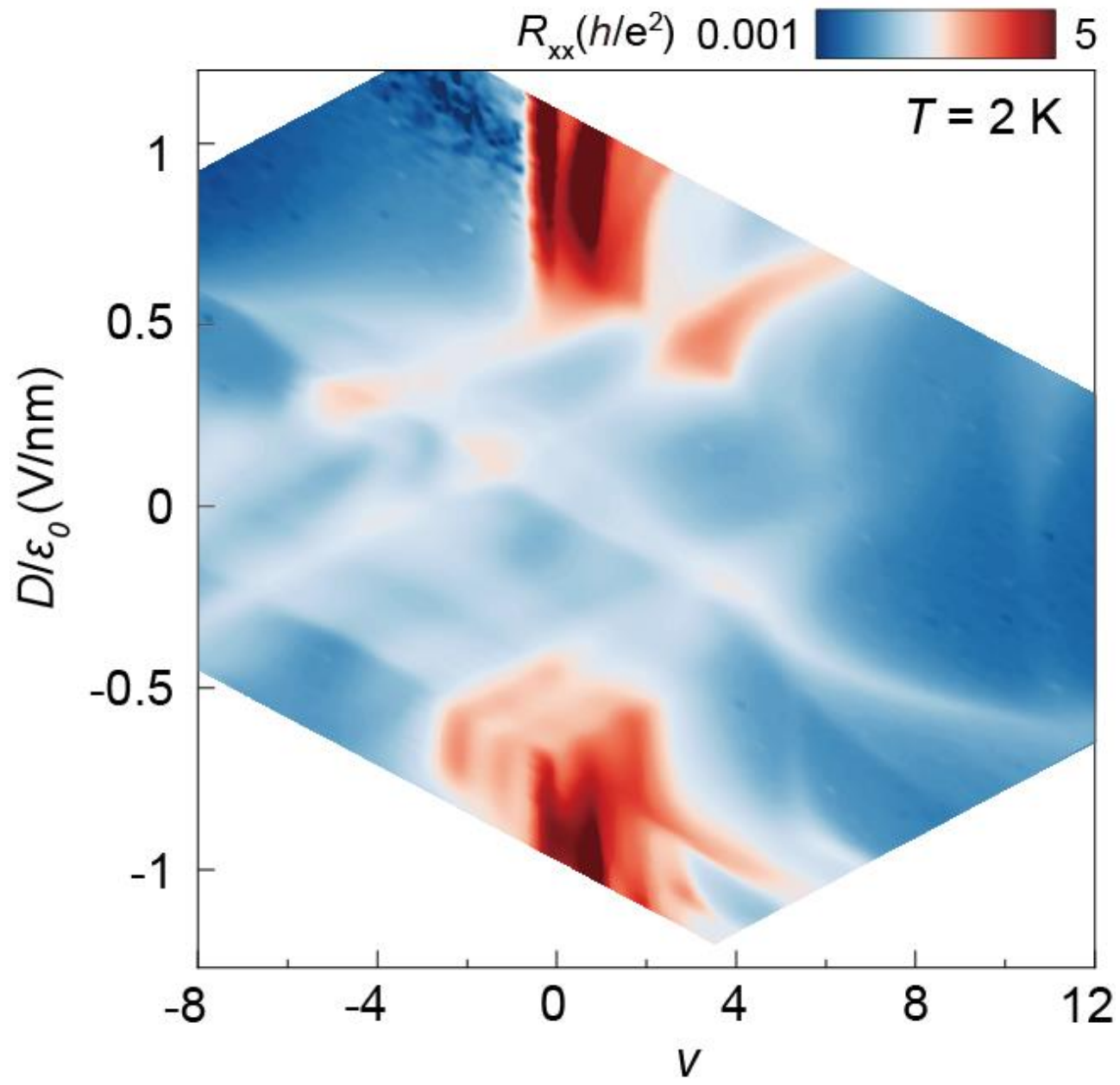

**Extended Data Figure 1 Phase diagram at 2 K.** For *D* in the range from -0.5 to 0.5 V/nm, resistive features are in parallel with top and bottom gates instead of *D*, representing electronic decoupling between top and bottom surface states. Further increasing *D*, one could see correlated insulators at integer fillings of moiré bands.

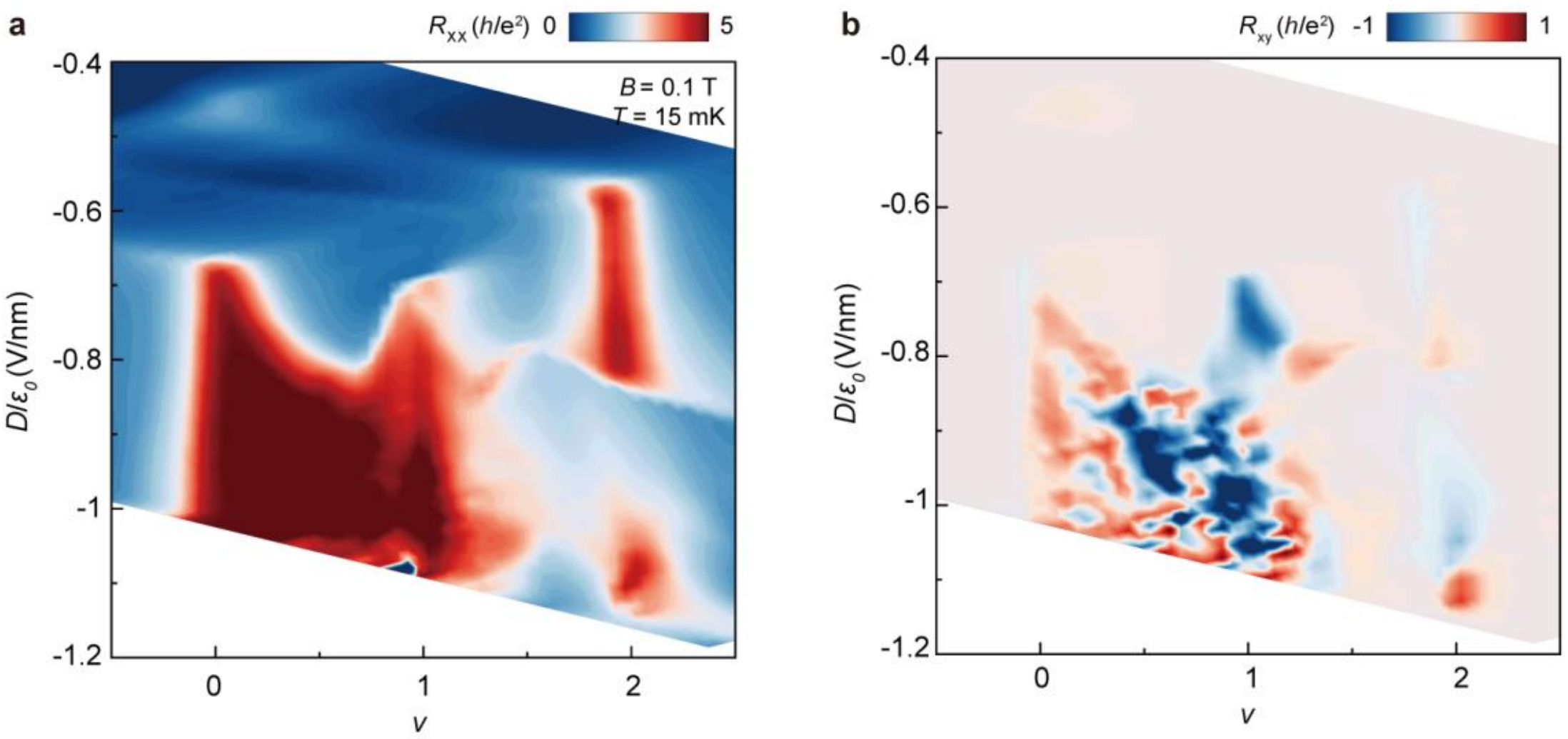

**Extended Data Figure 2 Phase diagram at *B*=0.1 T. a,** Longitudinal resistance at *T* = 15 mK. **b,** Hall resistance mapping, where a Chern insulator can be observed at *v* = 1 in the range from *D* = -0.7 to -0.8 V/nm.

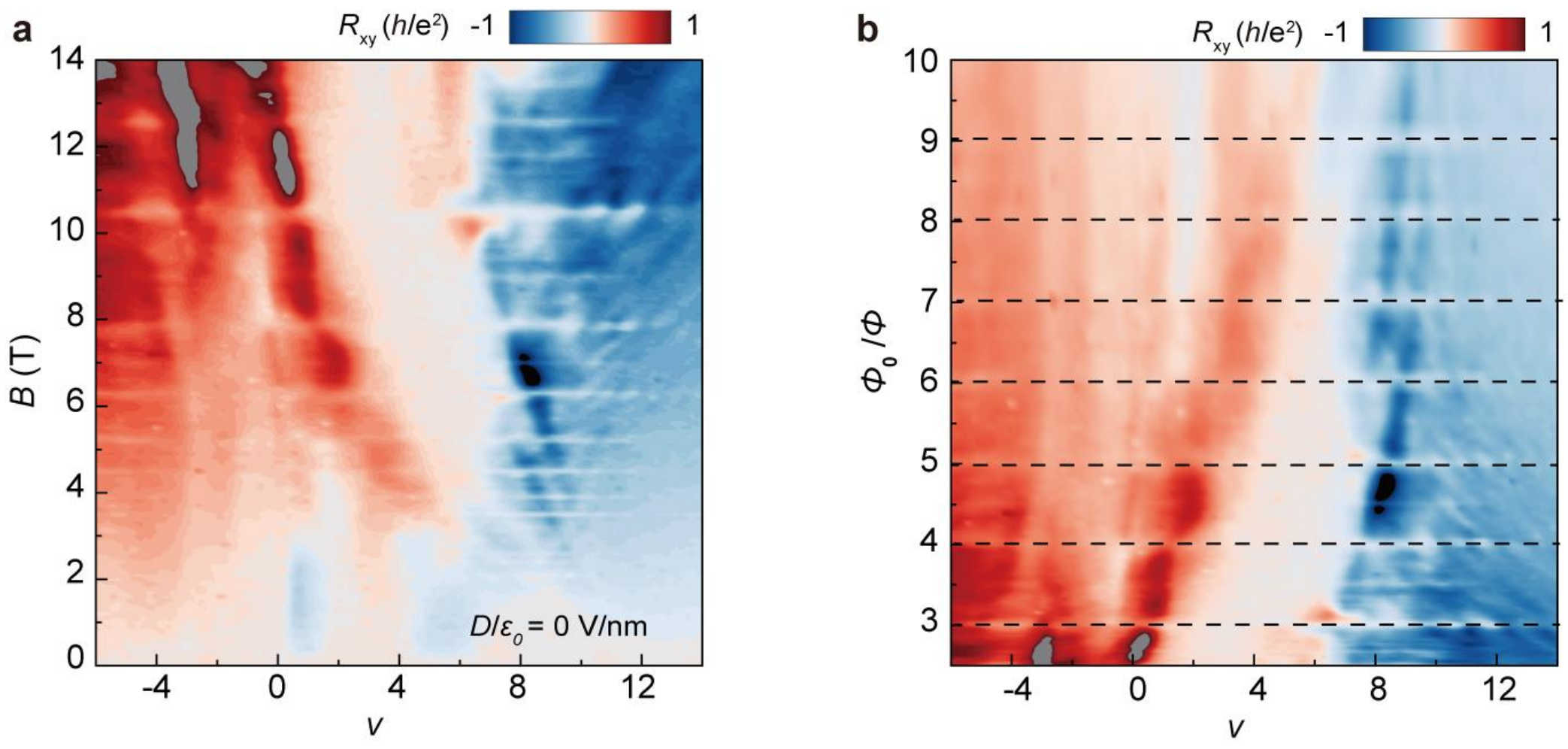

**Extended Data Figure 3 Brown-Zak oscillation observed at *D* = 0 V/nm.** The horizontal features independent of $v$ (**a**) are identified, and rearranged to be at integer index of the vertical axis (**b**) by using $\Phi = B * \frac{\sqrt{3}}{2}\lambda^2$ where $\lambda$ = 12.3 nm. The twisted angle between graphene and hBN thus is calculated as 0.67 degrees.

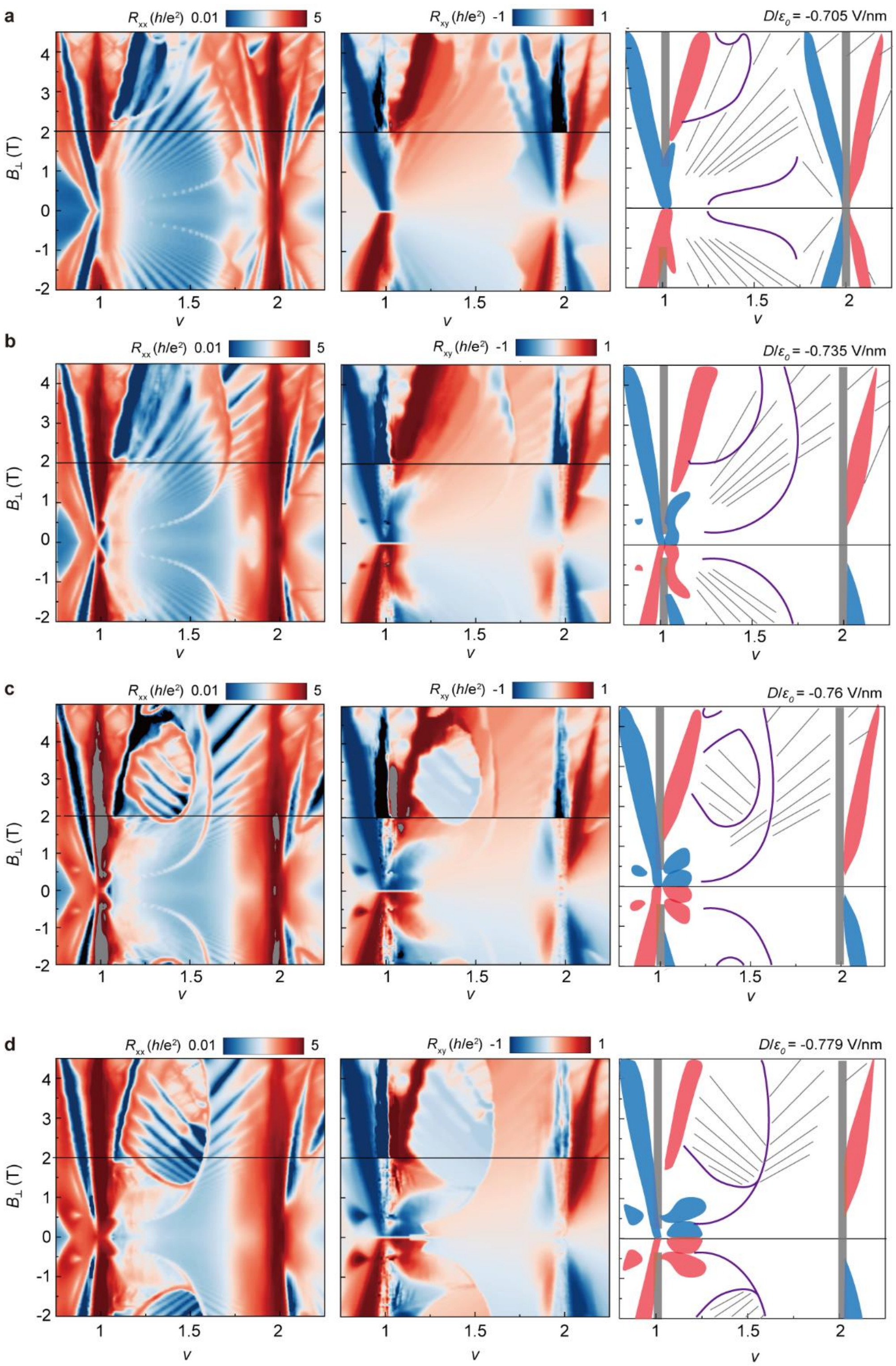

**Extended Data Figure 4 A full description of the evolution of Chern insulator at $\nu$ = 1 and Chern wings as a function of displacement fields.** Longitudinal resistance (left column), Hall resistance (middle) and schematics (right) are plotted as a function of $B$ (with the step of 0.05 T) at $D$ = -0.705 (**a**), -0.735 (**b**), -0.760 (**c**), -0.779 (**d**) V/nm. The data in the range of $B$ = -2 T and 2 T are symmetrized (longitudinal resistance) and anti-symmetrized (Hall resistance), separated by a black horizontal line from original data taken above 2 T. See color scheme of schematics in Figure 2.

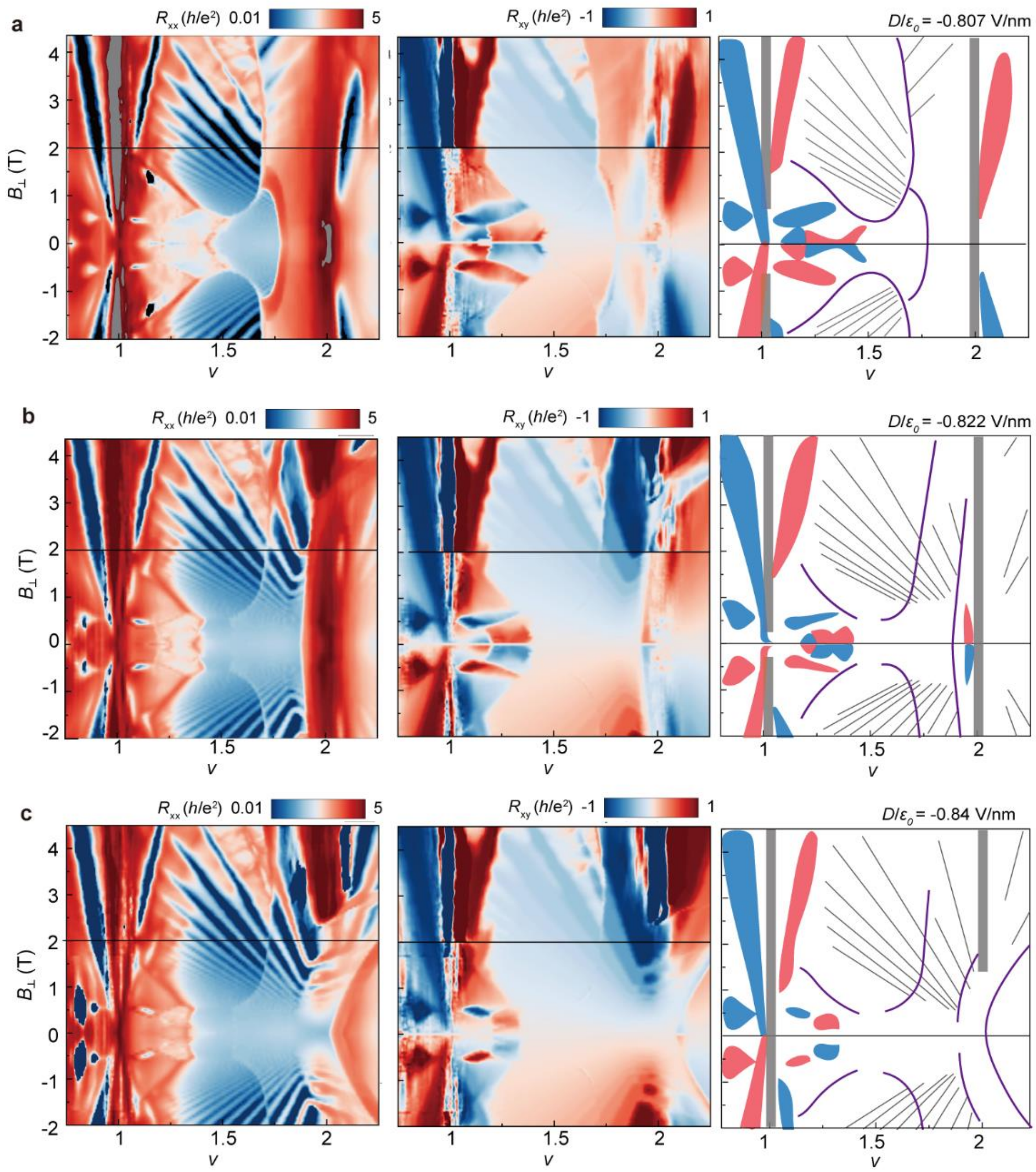

**Extended Data Figure 5 Continued evolution of Chern insulator at $\nu$ = 1 and Chern wings as a function of displacement fields.** Longitudinal resistance (left column), Hall resistance (middle) and schematics (right) are plotted as a function of *B* (with the step of 0.05 T) at *D* = -0.807 (**a**), -0.822 (**b** and -0.84 (**c**). The data in the range of *B* = -2 T and 2 T are symmetrized (longitudinal resistance) and anti-symmetrized (Hall resistance), separated by a black horizontal line from original data taken above 2 T. In the schematics, the color scheme is the same as that in Figure 2.

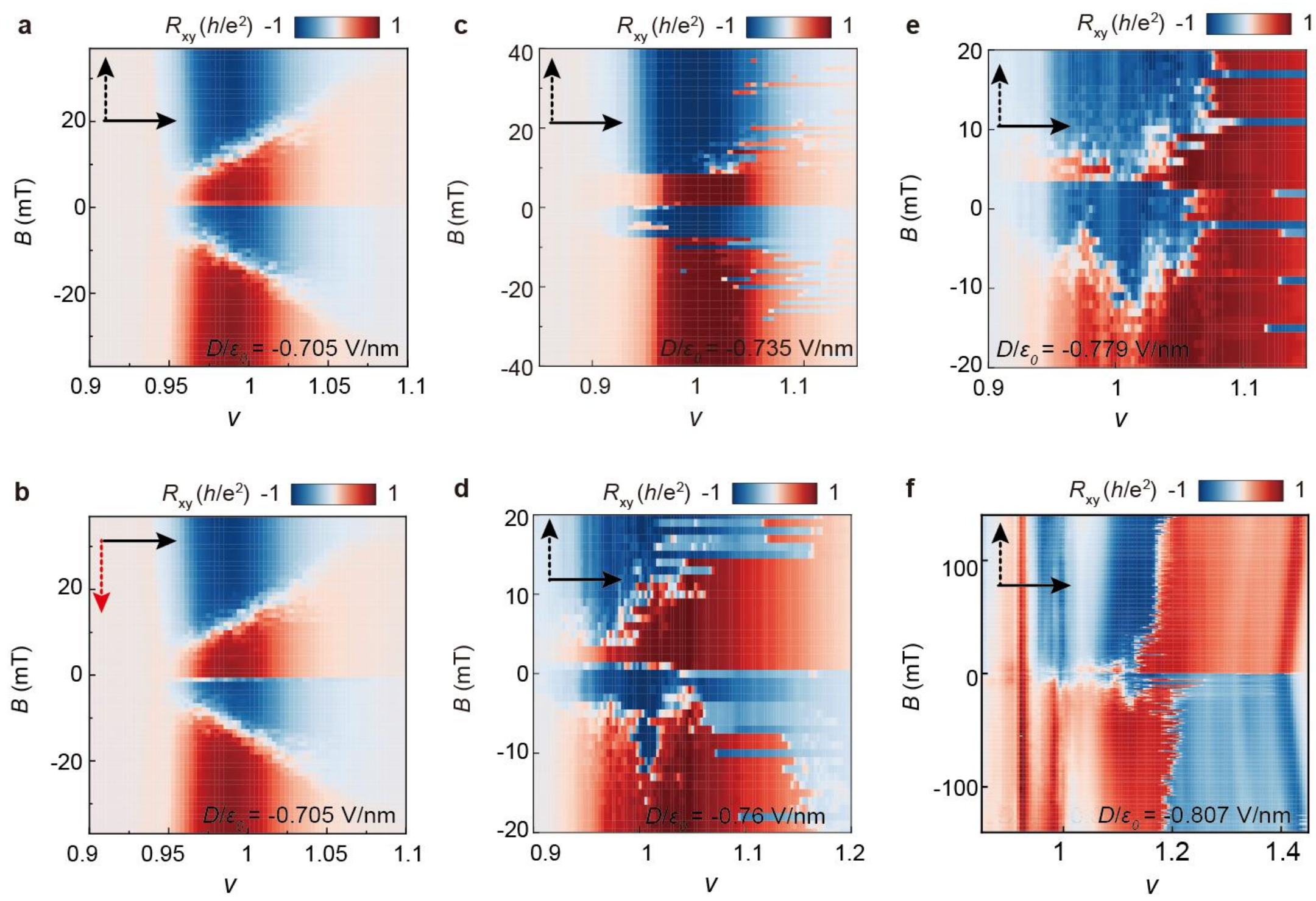

**Extended Data Figure 6 Switching orbital order driven by carrier density and magnetic field as a function of displacement field.** $D$ = -0.705 (**a, b**), -0.735 (**c**), -0.760 (**d**), -0.779 (**e**) and -0.807 (**f**) V/nm. There is an obvious difference between **a-b** and **c-f**: the phase boundaries between two orders in finite $B$ fields is relatively stable in **a-b**, whereas in **c-f** horizontal red and blue lines abruptly emerge in blue and red regimes, respectively. The latter indicates stochastic switching of orbital order during the transition driven by varying carrier density. The absence of such behaviors in panel **a-b** is associated with the diminishing hysteresis loop when sweeping carrier density.

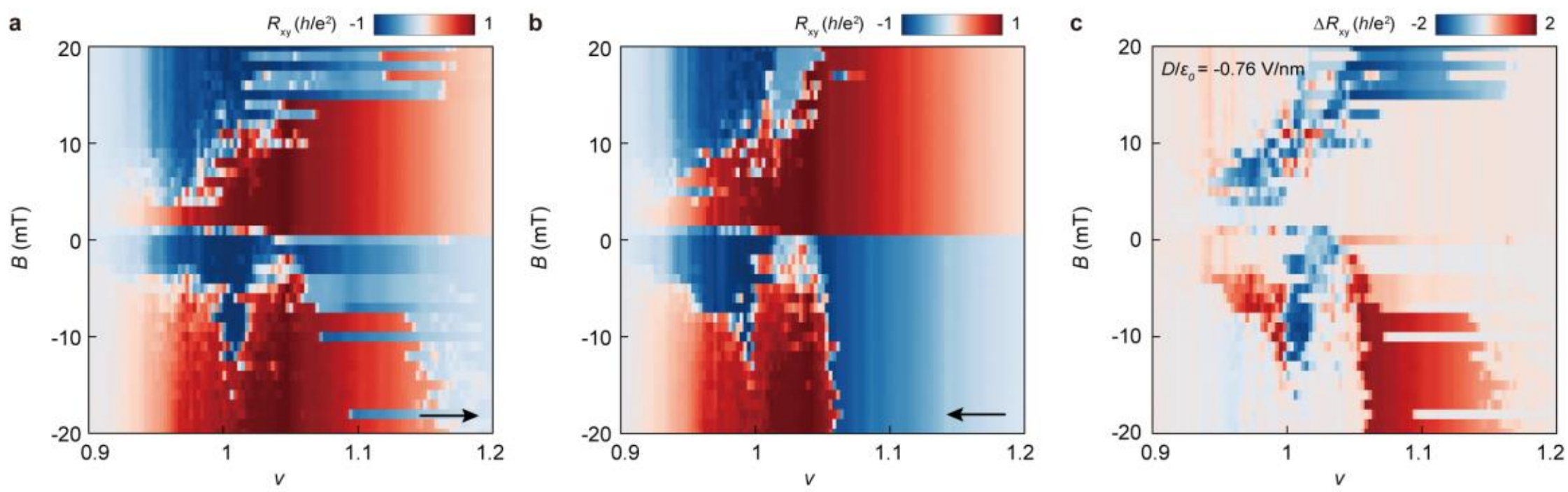

**Extended Data Figure 7 Electric switching of orbital orders at $D$ = -0.76 V/nm.** By subtracting the forwards sweeping of $v$ (**a**) from the backwards sweeping (**b**), the difference map is shown in **c**.

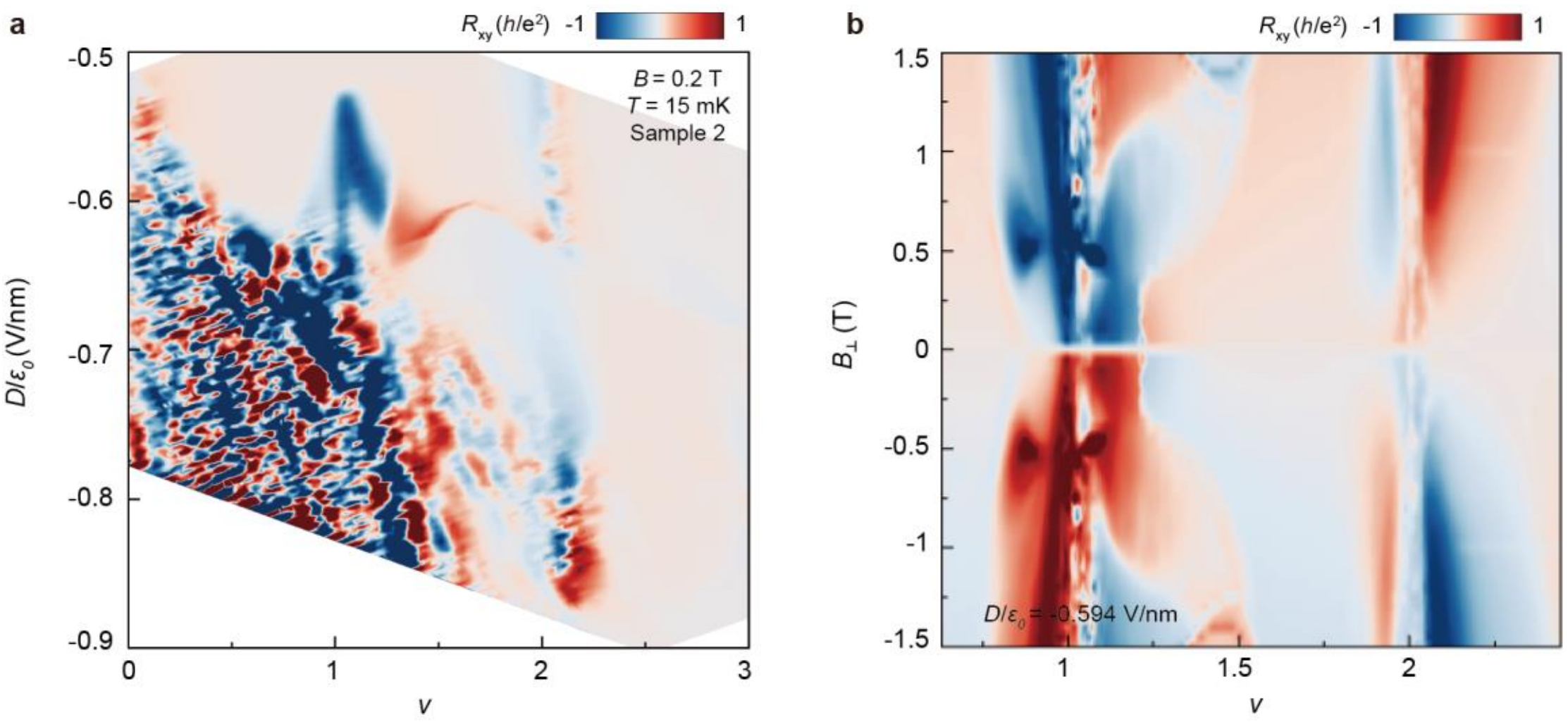

**Extended Data Figure 8 Additional device with moiré wavelength ~14.4 nm. a,** $D$-$\nu$ phase diagram at $B$ = 0.2 T. Besides the Chern insulator at $\nu$=1, a Chern wing emerges on the right side. **b,** At $D$ = -0.594 V/nm, Chern wings can be clearly identified.